\documentclass[twocolumn,aps,prd,superscriptaddress,floatfix,nofootinbib]{revtex4-2}
\usepackage[normalem]{ulem}
\usepackage{amsmath}
\usepackage[utf8]{inputenc}
\usepackage{amssymb}
\usepackage{gensymb}
\usepackage{bm}
\usepackage{graphicx}
\usepackage{float}
\usepackage{subcaption}
\usepackage[dvipsnames]{xcolor}
\usepackage{placeins}
\usepackage{braket}
\usepackage[colorlinks, linkcolor=blue, citecolor=blue, urlcolor=blue, breaklinks=red]{hyperref}
\usepackage{physics}
\usepackage{orcidlink}

\newcommand{\rh}{r_{+}}
\newcommand{\rmi}{r_{-}}
\newcommand{\Thw}{T_{H}}
\newcommand{\CI}{\mathcal{C}_{\mathrm{I}}}
\newcommand{\CII}{\mathcal{C}_{\mathrm{II}}}
\newcommand{\BI}{B_{\mathrm{I}}}
\newcommand{\BII}{B_{\mathrm{II}}}
\newcommand{\BmaxI}{\mathcal{B}_{\max}^{\mathrm{I}}}
\newcommand{\BmaxII}{\mathcal{B}_{\max}^{\mathrm{II}}}
\newcommand{\vrho}{\varrho}
\newcommand{\gc}{g_{c}}

\begin{document}

\title{Effect of the magnetic monopole charge on Dirac entanglement and Bell non-locality in Hayward spacetime}

\author{Abdessamie Chhieb~\!\!\orcidlink{0009-0007-0711-4747}}
\email[Corresponding author: ]{abdessamie.chhieb@ump.ac.ma}
\affiliation{Interdisciplinary Laboratory of Physics, Computer Science, and Oncology (LIPIO),\\ Mohamed First University, Oujda,  Morocco}
\affiliation{National Institute For Particle Physics and Applications (NIPPA),\\
Oujda, Morocco}
\author{Mohamed Ouchrif~\!\!\orcidlink{0000-0002-2954-1420}}
\email{m.ouchrif@ump.ac.ma}
\affiliation{Interdisciplinary Laboratory of Physics, Computer Science, and Oncology (LIPIO),\\ Mohamed First University, Oujda,  Morocco}
\affiliation{National Institute For Particle Physics and Applications (NIPPA),\\
Oujda, Morocco}
\begin{abstract}
We investigate the bipartite quantum correlations of Dirac fields in the background
of a Hayward regular black hole, in which the central singularity is replaced by a de
Sitter core characterised by the regularity parameter $g$. Two fermionic modes
initially prepared in a Bell-type entangled state are shared between an asymptotic
observer (Alice) and a static observer (Bob) hovering at fixed Schwarzschild-Hayward
coordinate near the event horizon. Through the standard Damour--Ruffini single-mode
Bogoliubov transformation, Bob's mode splits into a physically accessible exterior
mode $\BI$ and a physically inaccessible interior mode $\BII$. We derive in closed
form the reduced density matrices $\vrho_{A\BI}$ and $\vrho_{A\BII}$ and quantify
their quantum correlations through two complementary measures: the Wootters
concurrence $\mathcal{C}$ and the Clauser--Horne--Shimony--Holt (CHSH) Bell parameter
$\mathcal{B}_{\max}$. We find that (i)~the accessible correlations $\CI$, $\BmaxI$
never vanish even in the infinite-temperature limit, the Pauli exclusion principle
preventing total decoherence; (ii)~the inaccessible correlations $\CII$, $\BmaxII$
develop a non-zero value that witnesses the entanglement redistribution across the
horizon, with $\CII$ approaching $\CI$ as $\Thw\to\infty$; (iii)~the accessible
CHSH parameter violates the CHSH inequality at every finite, positive Hawking
temperature for $\alpha=\pi/4$, with $\BmaxI=2\sqrt{2/(\exp(-\omega/\Thw)+1)}$,
interpolating between the Tsirelson bound $2\sqrt{2}$ as $\Thw\to 0$ and the
classical limit $2$ as $\Thw\to\infty$; (iv)~the inaccessible CHSH parameter is
bounded above by the classical limit $2$ and therefore never violates Bell's
inequality. The regularity parameter $g$ enters all quantities only through the
Hayward Hawking temperature, which decreases monotonically as $g$ grows from $0$
to the extremal value $\gc=(16/27)^{1/3}M\simeq 0.8399\,M$. A direct comparison
with the bosonic (scalar-field) analogue, performed at matched values of the
Hawking parameter, makes the fermionic origin of our results explicit:
$\CI$ obeys the Pauli-protected lower bound $\sin(2\alpha)/\sqrt{2}$ and never
vanishes, whereas the scalar concurrence decoheres completely in the
infinite-temperature limit. The closed-form derivations suggest that
singularity resolution and quantum-information preservation are connected
through a single mechanism: the suppression of surface gravity at fixed
ADM mass, mediated by Fermi--Dirac statistics.
\end{abstract}


\maketitle

\section{Introduction}\label{sec:intro}

The black-hole information paradox~\cite{hawking1975particle,Hawking1976breakdown}
remains one of the deepest open problems at the interface of quantum mechanics
and gravity. A particularly fruitful avenue has been to study the fate of
quantum entanglement of relativistic quantum fields propagating in black-hole
backgrounds, since Hawking radiation degrades the quantum correlations measured
by an exterior observer~\cite{AlsingMilburn2003,FuentesSchuller2005,
adesso2007entanglement,wang2010projective,Xu2014probing,Xu2014multipartite,
He2015property,He2016measurement}. For Dirac fields in the Schwarzschild
background, a body of work~\cite{Xu2014probing,Xu2014multipartite,He2015property,
He2016measurement} has established that the Pauli exclusion principle prevents
the complete loss of quantum correlations at infinite Hawking temperature, in
sharp contrast with the bosonic case~\cite{wang2010quantum}. More recently, the
analysis has been extended to charged Reissner--Nordstr\"om (RN) black
holes~\cite{chhieb2024dirac}, where the electric charge modulates the Hawking
temperature and hence the rate of decoherence.

In this work, we extend the framework of relativistic quantum information to a
\emph{singularity-free} geometry, the Hayward regular black
hole~\cite{hayward2006formation}, whose central singularity is replaced by a de
Sitter core regulated by a length-scale parameter $g$. The motivation is
threefold. First, regular black
holes~\cite{bardeen1968,dymnikova1992vacuum,ayon1998regular,hayward2006formation,
frolov2016information} are the simplest non-trivial generalisation of the
Schwarzschild geometry that preserve the asymptotic structure while curing the
curvature singularity; they thus provide a controlled laboratory in which to
ask whether, and how, the absence of a singularity changes the
quantum-information landscape near the horizon. Second, the Hayward temperature
$\Thw(g)$ is strictly smaller than the Schwarzschild temperature at fixed mass,
and vanishes at the extremal value $\gc=(16/27)^{1/3}M\simeq 0.84\,M$, so a
clean monotonic prediction can be made: as $g$ increases, the rate of
Hawking-induced decoherence \emph{decreases}, and quantum correlations should
be more efficiently preserved. Third, the recent characterisation of inner-horizon
instabilities in regular black holes~\cite{carballorubio2018phenomenological,
carballorubio2022inner} has identified the smoothing of the central singularity
as a generic phenomenological feature of new physics at the horizon scale,
which makes the quantum-information signatures of $g$ a particularly clean
probe.

We focus on two complementary quantifiers. (i)~The Wootters
concurrence~\cite{wootters1998quantum} measures bipartite entanglement
algebraically, with a simple closed form for X-states. (ii)~The
Clauser--Horne--Shimony--Holt (CHSH) Bell
parameter~\cite{clauser1969proposed,horodecki1995violating} probes nonlocal
correlations through the violation of the CHSH inequality. The two measures
are complementary: a state can be entangled without violating any Bell
inequality, so the joint analysis reveals which fraction of the entanglement
survives as a Bell-nonlocal resource and which fraction is downgraded to a
hidden-variable-compatible correlation. The combination of an algebraic
quantifier (concurrence) and an inequality-based quantifier (CHSH) provides
a sharper picture of decoherence than any single quantity alone.

Our analysis follows the single-mode Damour--Ruffini Bogoliubov
convention~\cite{damour1976black,jing2004hawking} adopted in the reference
literature on Dirac fields in
Schwarzschild~\cite{AlsingMilburn2003,wang2010projective,Xu2014probing,
Xu2014multipartite,He2015property} and in the RN
extension~\cite{chhieb2024dirac}. The validity and limitations of this
approximation, in particular following the analyses
of~\cite{Bruschi2010unruh,MartinMartinez2010}.

The paper is organised as follows. Section~\ref{sec:geometry} introduces the
Hayward geometry and the associated Hawking temperature. Section~\ref{sec:vacuum}
reviews the vacuum structure of the Dirac field. Section~\ref{sec:model} sets
up the physical model, including a schematic representation of the setup in
Fig.~\ref{fig:schema}, and derives the reduced density matrices in both
bipartitions. Sections~\ref{sec:concurrence} and~\ref{sec:bell} compute the
concurrence and the CHSH parameter, respectively, with explicit limits and
Hayward dependence.
Section~\ref{sec:results} presents the numerical results and discusses, figure
by figure, the influence of the four control parameters $(g,M,\omega,\alpha)$.
Section~\ref{sec:conclusion} summarises the main findings and indicates
directions for future work. Throughout the paper we adopt natural units
$G=\hbar=c=k_{B}=1$.

\section{Hayward black hole geometry and Hawking temperature}\label{sec:geometry}

The line element of the Hayward regular black hole reads~\cite{hayward2006formation}
\begin{equation}
ds^{2}=-f(r)\,dt^{2}+f(r)^{-1}\,dr^{2}+r^{2}\,d\Omega^{2},
\label{eq:metric}
\end{equation}
with $d\Omega^{2}=d\theta^{2}+\sin^{2}\theta\,d\varphi^{2}$ and the lapse function
\begin{equation}
f(r)=1-\frac{2Mr^{2}}{r^{3}+2g^{3}},
\label{eq:lapse}
\end{equation}
where $M$ is the ADM mass and $g$ the regularity (Hayward) parameter. For
$g\to 0$ one recovers the Schwarzschild lapse $f(r)=1-2M/r$. Near the origin,
$f(r)\to 1-Mr^{2}/g^{3}$, which is the de Sitter form with effective
cosmological constant
\begin{equation}
\Lambda_{\mathrm{eff}}=\frac{3M}{g^{3}}.
\label{eq:Lambda_eff}
\end{equation}
The Kretschmann scalar is finite everywhere, signalling the absence of any
curvature singularity. The Hayward metric is sourced by a non-linear
electrodynamics or by an effective stress-energy tensor representing
quantum-gravity corrections~\cite{ayon1998regular,hayward2006formation}; the
exact origin of the source does not affect the propagation of probe Dirac
fields on this background.

The event horizon $\rh$ is the largest real root of $f(\rh)=0$,
\begin{equation}
\rh^{3}-2M\rh^{2}+2g^{3}=0.
\label{eq:horizon}
\end{equation}
Real positive roots exist only for $g\leq\gc$. The extremal condition imposes
Eq.~\eqref{eq:horizon} and its derivative $3\rh^{2}-4M\rh=0$ to hold
simultaneously, giving $\rh=4M/3$ and, after substitution,
\begin{equation}
\gc=\Bigl(\frac{16}{27}\Bigr)^{\!1/3}M\simeq 0.8399\,M.
\label{eq:gcrit}
\end{equation}
For $g>\gc$ no horizon exists and the metric describes a globally regular,
horizonless object. Throughout this work we restrict the analysis to the
physical regime $g\leq\gc$, where a regular black hole exists; we shall flag
the unphysical extrapolation explicitly whenever a formula is evaluated outside
this regime.

The Hawking temperature follows from the surface gravity
$\Thw=f'(\rh)/(4\pi)$. A direct computation of $f'(r)$ from
Eq.~\eqref{eq:lapse} gives
\begin{equation}
f'(r)=\frac{2Mr\,(r^{3}-4g^{3})}{(r^{3}+2g^{3})^{2}},
\label{eq:fprime}
\end{equation}
and hence
\begin{equation}
\boxed{\;\Thw=\frac{M\,\rh\,(\rh^{3}-4g^{3})}{2\pi\,(\rh^{3}+2g^{3})^{2}}.\;}
\label{eq:Hawking}
\end{equation}
For $g=0$, $\rh=2M$ and Eq.~\eqref{eq:Hawking} reduces to the Schwarzschild
result $\Thw=1/(8\pi M)$~\cite{hawking1975particle}. The temperature vanishes
at the extremal point $g=\gc$, where $\rh^{3}=4\gc^{3}$, signalling the
absence of Hawking emission for an extremal Hayward black hole.

It is instructive to compare~\eqref{eq:Hawking} with the analogous expression
for the Bardeen regular black hole~\cite{bardeen1968,ayon1998regular}, whose
lapse function $f_{\rm Bar}(r)=1-2Mr^{2}/(r^{2}+g^{2})^{3/2}$ leads to
\begin{equation}
\Thw^{\rm Bar}=\frac{M\,\rh\,(\rh^{2}-2g^{2})}{2\pi\,(\rh^{2}+g^{2})^{5/2}},
\quad \gc^{\rm Bar}=\frac{4M}{3\sqrt{3}}\simeq 0.7698\,M.
\label{eq:HawkingBar}
\end{equation}
Both geometries share the qualitative feature
$\Thw^{\rm RBH}(g)<\Thw^{\rm Schw}$ at fixed $M$ and $\rh$, with extremality
($\Thw\to 0$) reached at $g=\gc^{\rm RBH}$. The Bardeen formula is provided
here for completeness and will be referred to in Sec.~\ref{sec:bell} when
discussing the universal ordering of the CHSH parameter across regular
geometries.

\section{Vacuum structure of the Dirac field}\label{sec:vacuum}

In a generic curved background, the Dirac equation reads~\cite{brill1957interaction}
\begin{equation}
\bigl[\gamma^{a}e^{\mu}_{\;a}(\partial_{\mu}+\Gamma_{\mu})\bigr]\psi=0,
\label{eq:Dirac}
\end{equation}
with $e^{\mu}_{\;a}$ the inverse tetrad and $\Gamma_{\mu}$ the spin connection.
In the Hayward metric, separating variables in the tortoise coordinate
$r_{*}=\int dr/f(r)$ leads, in the standard
way~\cite{jing2004hawking,wang2010projective}, to positive-frequency outgoing
solutions outside and inside the event horizon:
\begin{align}
\psi^{\mathrm{I+}}_{k} &= \zeta\,e^{-i\omega u},\quad (r>\rh),\label{eq:psiI}\\
\psi^{\mathrm{II+}}_{k} &= \zeta\,e^{+i\omega u},\quad (r<\rh),\label{eq:psiII}
\end{align}
where $\zeta$ is a four-component Dirac spinor and $u=t-r_{*}$ the retarded
null coordinate. Following the Damour--Ruffini analytic
continuation~\cite{damour1976black} and the single-mode Bogoliubov treatment
used throughout the Dirac relativistic-quantum-information
literature~\cite{AlsingMilburn2003,wang2010projective,Xu2014probing,
Xu2014multipartite,He2015property,chhieb2024dirac}, the Kruskal vacuum and
the single-particle excited state of mode $k$ are related to the corresponding
Schwarzschild--Hayward Fock states by
\begin{align}
\ket{0}_{K} &= \frac{\ket{0}_{\mathrm{I}}\otimes\ket{0}_{\mathrm{II}}}{\sqrt{e^{-\omega/\Thw}+1}}
+ \frac{\ket{1}_{\mathrm{I}}\otimes\ket{1}_{\mathrm{II}}}{\sqrt{e^{\omega/\Thw}+1}},
\label{eq:vac}\\[2pt]
\ket{1}_{K} &= \ket{1}_{\mathrm{I}}\otimes\ket{0}_{\mathrm{II}}.
\label{eq:exc}
\end{align}
Here, the state $\ket{n}_{\mathrm{I}}$ ($\ket{n}_{\mathrm{II}}$) is the
$n$-particle excitation of the outgoing mode in region $\mathrm{I}$
($\mathrm{II}$). The vacuum~\eqref{eq:vac} is the unique vacuum
seen by a static observer hovering at fixed coordinate $r=\rh+\epsilon$ outside
the horizon, in the Boulware--Hartle--Hawking
sense~\cite{unruh1976notes,israel1976thermo,hartle1976path}; we elaborate on
this in Sec.~\ref{sec:limitations}.

The Pauli exclusion principle restricts Eq.~\eqref{eq:vac} to only two
terms, in contrast with the infinite geometric sum that arises for bosonic
fields~\cite{birrell1984quantum,wang2010quantum}. Explicitly, for a scalar
(spin-$0$) field the Bogoliubov transformation produces an
infinite-dimensional Fock decomposition,
\begin{equation}
\ket{0}^{\mathrm{bos}}_{K}=\frac{1}{\cosh r_{\omega}}
\sum_{n=0}^{\infty}(\tanh r_{\omega})^{n}\,\ket{n}_{\mathrm{I}}\ket{n}_{\mathrm{II}},
\label{eq:vac-bos}
\end{equation}
whereas Pauli exclusion truncates the Dirac sector to the two-level
decomposition of Eq.~\eqref{eq:vac}~\cite{Alsing2006,martin2011fermionic}.
This truncation is physically significant: it bounds the entanglement
entropy and modifies the concurrence structure, so that the maximal
entanglement degradation for fermionic modes remains \emph{bounded},
whereas bosonic modes can undergo complete decoherence in the
infinite-temperature limit. Equivalently, the anticommutation relations
of the fermionic field operators impose the unitarity constraint
$|\alpha_{\omega}|^{2}+|\beta_{\omega}|^{2}=1$ on the Bogoliubov
coefficients, in contrast with the bosonic relation
$|\alpha_{\omega}|^{2}-|\beta_{\omega}|^{2}=1$~\cite{Takagi1986}; this
constraint is responsible for the X-type form of the reduced density
matrices $\vrho_{A\BI}$ and $\vrho_{A\BII}$ derived below, and has no
bosonic analogue. The thermal occupation that the exterior observer
ultimately reads off therefore carries a \emph{Fermi--Dirac} factor
$(e^{\omega/\Thw}+1)^{-1}$, whereas the bosonic counterpart would be the
\emph{Bose--Einstein} factor $(e^{\omega/\Thw}-1)^{-1}$ that diverges
as $\Thw\to\infty$. The structural lower bound on the accessible
concurrence reported in Eq.~\eqref{eq:CIbound} below is a direct
consequence of this difference.

We introduce the convenient abbreviations
\begin{equation}
p\equiv\frac{1}{e^{-\omega/\Thw}+1},
\quad
q\equiv\frac{1}{e^{\omega/\Thw}+1},
\quad p+q=1,
\label{eq:pq}
\end{equation}
so that Eq.~\eqref{eq:vac} takes the compact form
\begin{equation}
\ket{0}_{K}=\sqrt{p}\,\ket{0}_{\mathrm{I}}\ket{0}_{\mathrm{II}}+\sqrt{q}\,\ket{1}_{\mathrm{I}}\ket{1}_{\mathrm{II}}.
\label{eq:vac-short}
\end{equation}
The Hayward specificity enters only through $\Thw=\Thw(g,M,\rh)$ via
Eq.~\eqref{eq:Hawking}.

\section{Physical model and reduced density matrices}\label{sec:model}

\begin{figure*}[!t]
\centering
\includegraphics[width=0.86\textwidth]{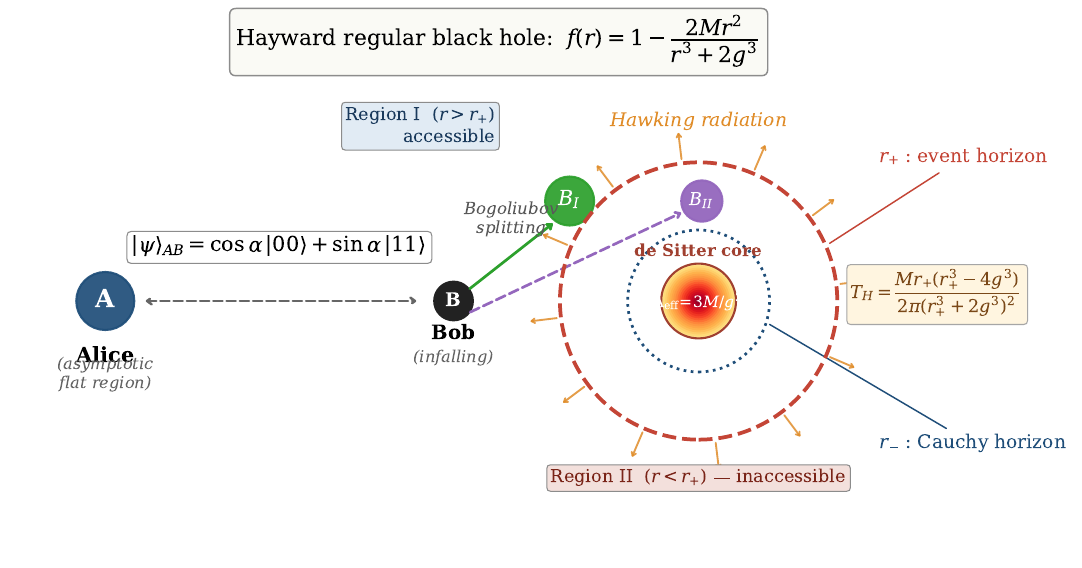}
\caption{Schematic representation of the physical setup. Alice, with qubit
$A$, stays in the asymptotically flat region. Bob, with qubit $B$, hovers
statically at fixed Schwarzschild--Hayward coordinate $r=\rh+\epsilon$ near
the event horizon of a Hayward regular black hole whose central singularity
is replaced by a de Sitter core (orange) of effective cosmological constant
$\Lambda_{\mathrm{eff}}=3M/g^{3}$. The metric admits an inner Cauchy horizon
$\rmi$ (blue dotted) and an event horizon $\rh$ (red dashed). At $\rh$,
Bob's mode splits via the Damour--Ruffini Bogoliubov transformation into an
exterior mode $\BI$ (accessible) and an interior mode $\BII$ (causally hidden
behind $\rh$). Hawking radiation, sourced by the surface gravity at $\rh$
with temperature $\Thw$, decoheres the initial Bell state $\ket{\psi}_{AB}$.}
\label{fig:schema}
\end{figure*}

\subsection{Setup}\label{sec:setup}
We consider two fermionic modes $A$ and $B$ initially prepared in a
Bell-type entangled state
\begin{equation}
\ket{\psi}_{AB}=\cos\alpha\,\ket{0}_{A}\ket{0}_{B}+\sin\alpha\,\ket{1}_{A}\ket{1}_{B},
\label{eq:initial}
\end{equation}
where $\alpha\in[0,\pi/2]$ controls the initial entanglement; $\alpha=0,\pi/2$
correspond to product states and $\alpha=\pi/4$ to the maximally entangled
Bell state. For brevity we write $a\equiv\cos\alpha$, $b\equiv\sin\alpha$.
Alice carries mode $A$ and remains in the asymptotically flat region; Bob
carries mode $B$ and is a \emph{static observer} at fixed coordinate
$r=\rh+\epsilon$ outside the event horizon, so that his accelerated
worldline samples the Hartle--Hawking vacuum. (A free-falling observer would
see no Hawking quanta in the equivalent-principle sense; the splitting
\eqref{eq:vac} pertains to the static observer.) The Hawking effect splits
Bob's Kruskal mode into an exterior (accessible) component $\BI$ and an
interior (inaccessible) component $\BII$, as depicted in
Fig.~\ref{fig:schema}.

\subsection{Tripartite state in Hayward modes}
Substituting Eqs.~\eqref{eq:exc} and~\eqref{eq:vac-short} into
\eqref{eq:initial}, with Alice's mode untouched, we expand
$\ket{0}_{B}\to\sqrt{p}\,\ket{0}_{\BI}\ket{0}_{\BII}+\sqrt{q}\,\ket{1}_{\BI}\ket{1}_{\BII}$
and $\ket{1}_{B}\to\ket{1}_{\BI}\ket{0}_{\BII}$. The initial state becomes the
tripartite pure state
\begin{equation}
\ket{\psi}_{A\BI\BII}=a\sqrt{p}\,\ket{000}+a\sqrt{q}\,\ket{011}+b\,\ket{110}.
\label{eq:tripartite}
\end{equation}
Normalisation is preserved: $a^{2}p+a^{2}q+b^{2}=a^{2}+b^{2}=1$.

\subsection{Reduced state $\vrho_{A\BI}$ (accessible)}
Since $\BII$ is causally hidden behind $\rh$ for the static external observer,
we must trace it out. A term $\ket{\phi_{i}}\!\bra{\phi_{j}}$ of
$\ket{\psi}\!\bra{\psi}$ survives the partial trace over $\BII$ if and only
if the two $\BII$ labels coincide. Of the nine cross-terms
of~\eqref{eq:tripartite}, five survive: the three diagonal terms plus the
two cross-terms between $\ket{000}$ and $\ket{110}$, since both have
$\BII=0$. Collecting them in the computational basis
$\{\ket{00},\ket{01},\ket{10},\ket{11}\}$ for $(A,\BI)$,
\begin{equation}
\vrho_{A\BI}=
\begin{pmatrix}
a^{2}p & 0 & 0 & ab\sqrt{p}\\
0 & a^{2}q & 0 & 0\\
0 & 0 & 0 & 0\\
ab\sqrt{p} & 0 & 0 & b^{2}
\end{pmatrix}.
\label{eq:rhoABI}
\end{equation}
This is an X-type density matrix. Trace and positivity are immediate:
$\mathrm{tr}\,\vrho_{A\BI}=a^{2}(p+q)+b^{2}=1$, and the spectrum is
$\{a^{2}p+b^{2},a^{2}q,0,0\}\succeq 0$.

\subsection{Reduced state $\vrho_{A\BII}$ (inaccessible)}
An analogous calculation, this time tracing over $\BI$, retains the diagonal
terms plus the cross-terms between $\ket{011}$ and $\ket{110}$ (both have
$\BI=1$). The result reads
\begin{equation}
\vrho_{A\BII}=
\begin{pmatrix}
a^{2}p & 0 & 0 & 0\\
0 & a^{2}q & ab\sqrt{q} & 0\\
0 & ab\sqrt{q} & b^{2} & 0\\
0 & 0 & 0 & 0
\end{pmatrix}.
\label{eq:rhoABII}
\end{equation}
The spectrum is $\{a^{2}p,a^{2}q+b^{2},0,0\}$. Both
Eqs.~\eqref{eq:rhoABI} and~\eqref{eq:rhoABII} are X-shaped in the
computational basis, which dramatically simplifies the evaluation of all three
quantum-information measures used below.

\section{Wootters concurrence}\label{sec:concurrence}
For an arbitrary two-qubit state $\vrho$, the Wootters
concurrence~\cite{wootters1998quantum} is
\begin{equation}
C(\vrho)=\max\bigl\{0,\sqrt{\lambda_{1}}-\sqrt{\lambda_{2}}-\sqrt{\lambda_{3}}-\sqrt{\lambda_{4}}\bigr\},
\label{eq:Wootters}
\end{equation}
with $\lambda_{i}$ the eigenvalues, in decreasing order, of
$R=\vrho\,(\sigma_{y}\otimes\sigma_{y})\,\vrho^{*}\,(\sigma_{y}\otimes\sigma_{y})$.
For an X-state with elements $\vrho_{ij}$,
Eq.~\eqref{eq:Wootters} reduces to~\cite{hashemi2012genuinely}
\begin{equation}
C(\vrho_{X})=2\max\bigl\{0,|\vrho_{14}|-\sqrt{\vrho_{22}\vrho_{33}},|\vrho_{23}|-\sqrt{\vrho_{11}\vrho_{44}}\bigr\}.
\label{eq:C-X}
\end{equation}
From Eq.~\eqref{eq:rhoABI}: $\vrho_{14}=ab\sqrt{p}$, $\vrho_{23}=0$,
$\vrho_{22}\vrho_{33}=0$, and $\vrho_{11}\vrho_{44}=a^{2}b^{2}p$, whence
\begin{equation}
\boxed{\;\CI=C(\vrho_{A\BI})=2ab\sqrt{p}=\frac{\sin 2\alpha}{\sqrt{e^{-\omega/\Thw}+1}}.\;}
\label{eq:CI}
\end{equation}
Similarly, from Eq.~\eqref{eq:rhoABII}: $\vrho_{14}=0$, $\vrho_{23}=ab\sqrt{q}$,
$\vrho_{11}\vrho_{44}=0$, $\vrho_{22}\vrho_{33}=a^{2}b^{2}q$, so
\begin{equation}
\boxed{\;\CII=C(\vrho_{A\BII})=2ab\sqrt{q}=\frac{\sin 2\alpha}{\sqrt{e^{\omega/\Thw}+1}}.\;}
\label{eq:CII}
\end{equation}

\emph{Limits.} As $\Thw\to 0$: $p\to 1$, $q\to 0$, $\CI\to\sin 2\alpha$
(initial Bell entanglement preserved in the accessible sector), $\CII\to 0$
(no entanglement leaks to the interior). As $\Thw\to\infty$:
$p,q\to 1/2$, both $\CI$ and $\CII$ tend to $\sin(2\alpha)/\sqrt{2}>0$, a
fermionic protection against total entanglement loss due to Pauli
exclusion~\cite{Xu2014probing,He2015property,chhieb2024dirac}, witnessing
the symmetric redistribution of entanglement between the two sectors.
This is in sharp contrast with scalar (bosonic) fields, for which the
infinite geometric sum in the Bogoliubov decomposition leads to $\CI\to 0$
as $\Thw\to\infty$~\cite{wang2010quantum,chhieb2024dirac}, i.e.\ complete
decoherence. The Pauli exclusion principle thus provides a structural lower
bound $\CI\geq\sin(2\alpha)/\sqrt{2}$ that has no bosonic analogue.

\emph{Hayward dependence.} Both \eqref{eq:CI} and \eqref{eq:CII} depend on $g$
only through $\Thw=\Thw(g,M,\rh)$. Since $\Thw$ decreases monotonically with
$g$ in the physical regime $g\leq\gc$,
\begin{equation}
\frac{\partial \CI}{\partial g}\geq 0,\qquad
\frac{\partial \CII}{\partial g}\leq 0,
\label{eq:monotonicity}
\end{equation}
at fixed $M$ and $\omega$. The regularity parameter therefore enhances the
entanglement in the exterior and suppresses that leaking to the interior;
at extremality $g=\gc$, the initial Bell state is fully preserved
($\CI=\sin 2\alpha$, $\CII=0$).

\emph{Analytic structure and non-monotonic behaviour.}
The Hayward Hawking temperature $\Thw(g,M,\rh)$ of Eq.~\eqref{eq:Hawking}
vanishes at the extremal locus $\rh=4^{1/3}g$ and as $\rh\to\infty$, and
reaches an interior maximum $\Thw^{\max}=\Thw^{\max}(g,M)$ at an
intermediate radius $\rh^{*}=\rh^{*}(g,M)$ controlled by the
mass-regularity ratio. Because
$\CI=\sin(2\alpha)/\sqrt{1+e^{-\omega/\Thw}}$ is a monotonically
\emph{decreasing} function of $\Thw$ at fixed $\omega$ and $\alpha$, the
concurrence inherits a non-monotonic dependence on $\rh$:
\begin{itemize}
\item at $\rh=4^{1/3}g$ (extremal limit): $\Thw=0$,
$e^{-\omega/\Thw}\to 0$, and $\CI\to\sin(2\alpha)$ (maximal preservation);
\item at $\rh=\rh^{*}(g,M)$ (maximum temperature): $\CI$ reaches its
minimum, corresponding to the strongest Hawking effect;
\item as $\rh\to\infty$: $\Thw\to 0$ again, and $\CI$ recovers towards
$\sin(2\alpha)$.
\end{itemize}
This non-monotonic behaviour, sharply visible in Figs.~\ref{fig:1}--\ref{fig:2}
below, has no analogue in the Schwarzschild case ($g=0$), where
$\Thw=1/(4\pi\rh)$ is monotonically decreasing and $\CI$ monotonically
increases with $\rh$~\cite{PanYu2008}. The Hayward regularity scale
therefore introduces a qualitatively new feature in the entanglement
landscape, parallel to that observed in the
Reissner--Nordstr\"om~\cite{chhieb2024dirac} and dilatonic~\cite{He2016measurement}
contexts but with the characteristic Hayward signature
$\rh^{(\mathrm{ext})}=4^{1/3}g\simeq 1.587\,g$.

The transition thresholds at which $\CI$ drops below a chosen level
$\varepsilon$ follow analytically from
$\Thw(\rh^{\dagger},g,M)=\omega/\ln(1/\varepsilon^{2}-1)^{-1}$, giving
$\rh^{\dagger}\simeq 4^{1/3}g\,(1+\delta_{\omega})$ with a small,
$\omega$-dependent correction $\delta_{\omega}$ that scales linearly with
$g$. This linear scaling underlies the regular spacing of the
charge-induced transitions of Fig.~\ref{fig:1}, in agreement with the
numerical curves.

\emph{Frequency limits.}
In the high-frequency limit $\omega\to\infty$ one has $p\to 1$ and
$q\to 0$, yielding $\CI\to\sin(2\alpha)$ (maximal preservation) and
$\CII\to 0$. Conversely, in the low-frequency limit $\omega\to 0$ at fixed
$\Thw>0$, both $p$ and $q$ tend to $1/2$, leading to the symmetric
plateau $\CI=\CII=\sin(2\alpha)/\sqrt{2}$. These two analytical limits
explain quantitatively both the low-frequency plateau and the
high-frequency enhancement that will be seen in Fig.~\ref{fig:3} below,
in line with the dilatonic~\cite{Martin-Martinez2010} and the
RN~\cite{chhieb2024dirac} analyses.

\emph{Fermionic lower bound.}
The plateau value $\sin(2\alpha)/\sqrt{2}$ is the universal lower bound
of the accessible fermionic concurrence in the present setting:
\begin{equation}
\boxed{\;\CI\;\geq\;\frac{\sin 2\alpha}{\sqrt{2}}\qquad\text{(for any }\Thw>0\text{)}.\;}
\label{eq:CIbound}
\end{equation}
This bound is a direct consequence of the Fermi--Dirac statistics
discussed below Eq.~\eqref{eq:vac-bos}: the $+1$ in the denominator of
$\CI=\sin(2\alpha)/\sqrt{1+e^{-\omega/\Thw}}$ enforces a maximum of $2$
in the Fermi--Dirac factor, whereas the analogous Bose--Einstein factor
$(e^{\omega/\Thw}-1)^{-1}$ diverges as $\Thw\to\infty$ and can lead to
complete decoherence. The fermionic concurrence therefore never vanishes
completely regardless of the values of $g$, $\omega$, or $\rh$ in the
physical Hayward range, a feature that we will visualise in
Fig.~\ref{fig:6} through a direct comparison with the scalar-field
analogue.

\section{Bell--CHSH non-locality}\label{sec:bell}

The CHSH operator~\cite{clauser1969proposed,horodecki1995violating} is
\begin{multline}
\mathcal{B}_{\mathrm{CHSH}}=\bm{a}\cdot\bm{\sigma}\otimes(\bm{b}+\bm{b}')\cdot\bm{\sigma}\\
+\bm{a}'\cdot\bm{\sigma}\otimes(\bm{b}-\bm{b}')\cdot\bm{\sigma},
\label{eq:CHSH}
\end{multline}
with $\bm{a},\bm{a}',\bm{b},\bm{b}'$ unit vectors of $\mathbb{R}^{3}$, and
the CHSH inequality is $|\!\langle\mathcal{B}_{\mathrm{CHSH}}\rangle\!|\leq 2$.
The Tsirelson bound~\cite{tsirelson1980quantum} sets the quantum maximum at
$2\sqrt{2}$. For a generic two-qubit state $\vrho$, the
Horodecki criterion gives
\begin{equation}
\mathcal{B}_{\max}(\vrho)=2\sqrt{M(\vrho)},\quad M(\vrho)=t_{i}^{2}+t_{j}^{2},
\label{eq:Bmax}
\end{equation}
where $t_{i}^{2},t_{j}^{2}$ are the two largest of $T_{11}^{2},T_{22}^{2},T_{33}^{2}$,
and $T_{ii}=\mathrm{tr}[\vrho\,\sigma_{i}\otimes\sigma_{i}]$ are the diagonal
components of the correlation tensor. For an X-state with real off-diagonals,
\begin{align}
T_{11}&=2(\vrho_{14}+\vrho_{23}),\nonumber\\
T_{22}&=2(\vrho_{23}-\vrho_{14}),\label{eq:Tii-X}\\
T_{33}&=\vrho_{11}-\vrho_{22}-\vrho_{33}+\vrho_{44}.\nonumber
\end{align}

\subsection{Accessible CHSH parameter}
From Eq.~\eqref{eq:rhoABI}:
$T_{11}^{\mathrm{I}}=-T_{22}^{\mathrm{I}}=\sin 2\alpha\,\sqrt{p}$,
$T_{33}^{\mathrm{I}}=1-2a^{2}q$. For $\alpha=\pi/4$,
$(T_{11}^{\mathrm{I}})^{2}=(T_{22}^{\mathrm{I}})^{2}=p$,
$(T_{33}^{\mathrm{I}})^{2}=p^{2}\leq p$, so the two largest squared
components are $\{p,p\}$ and
\begin{equation}
\boxed{\;\BmaxI\big|_{\alpha=\pi/4}=2\sqrt{2p}=\frac{2\sqrt{2}}{\sqrt{e^{-\omega/\Thw}+1}}.\;}
\label{eq:BI-pi4}
\end{equation}

\emph{Limits and Bell violation.} (i)~$\Thw\to 0$: $p\to 1$,
$\BmaxI\to 2\sqrt{2}$, the Tsirelson bound. (ii)~$\Thw\to\infty$: $p\to 1/2$,
$\BmaxI\to 2$, the classical limit. (iii)~Bell violation requires
$\BmaxI>2$, i.e.\ $p>1/2$, which holds if and only if $\Thw>0$
(since $e^{-\omega/\Thw}<1$ for any finite positive $\omega$ and $\Thw$).
Consequently, the accessible state at
$\alpha=\pi/4$ violates the CHSH inequality at every finite, strictly positive
Hawking temperature, i.e.\ for all Hayward black holes with $g<\gc$.

For arbitrary $\alpha$,
\begin{equation}
M(\vrho_{A\BI})=p\sin^{2}2\alpha+\max\bigl\{p\sin^{2}2\alpha,(1-2a^{2}q)^{2}\bigr\}.
\label{eq:M-I}
\end{equation}

\subsection{Inaccessible CHSH parameter}
From Eq.~\eqref{eq:rhoABII}:
$T_{11}^{\mathrm{II}}=T_{22}^{\mathrm{II}}=\sin 2\alpha\,\sqrt{q}$,
$T_{33}^{\mathrm{II}}=2a^{2}p-1$. At $\alpha=\pi/4$, $a^{2}=1/2$,
$T_{33}^{\mathrm{II}}=p-1=-q$, so
\begin{equation}
\boxed{\;\BmaxII\big|_{\alpha=\pi/4}=2\sqrt{2q}=\frac{2\sqrt{2}}{\sqrt{e^{\omega/\Thw}+1}}\leq 2.\;}
\label{eq:BII-pi4}
\end{equation}
Violation would require $q>1/2$, impossible since $q\in(0,1/2)$. The
inaccessible state never violates CHSH, in agreement with the corresponding
results for Schwarzschild~\cite{Xu2014probing,He2015property} and
RN~\cite{chhieb2024dirac}.

\subsection{Universal ordering at fixed $M$}
As shown in Sec.~\ref{sec:geometry}, both the Hayward and Bardeen geometries
have $\Thw^{\rm RBH}\leq\Thw^{\rm Schw}$ at fixed $M$ and $\rh$, with
extremality reached at $\gc^{\rm Hay}\simeq 0.840\,M$ and
$\gc^{\rm Bar}\simeq 0.770\,M$. The CHSH parameter being a monotonically
\emph{decreasing} function of $\Thw$ in the accessible sector, this implies
\begin{equation}
\BmaxI\bigr|_{\mathrm{Schw}}\leq\BmaxI\bigr|_{\mathrm{Bar}}\leq\BmaxI\bigr|_{\mathrm{Hay}}
\quad\text{at fixed } M,\rh,\omega.
\label{eq:hierarchy}
\end{equation}
This ordering, which is a strict mathematical consequence of the explicit
forms~\eqref{eq:Hawking}--\eqref{eq:HawkingBar} and of the Pauli-protected
single-mode structure, formalises the qualitative statement that
\emph{regular geometries preserve quantum non-locality more efficiently than
their singular Schwarzschild counterpart at fixed ADM mass}. We stress that
the comparison is performed at a common value of the Schwarzschild radial
coordinate $\rh$; for each geometry, $\rh$ is related to $M$ and $g$ through
the respective horizon equation, so the physical interpretation of $\rh$
as an areal radius is shared, but the internal structure of the geometry
(de~Sitter core versus singularity) differs.

\section{Numerical results and discussion}\label{sec:results}

Before turning to the full parameter scans, we collect in
Table~\ref{tab:limits} the analytical limiting values of all six
bipartite measures at $\alpha=\pi/4$, for the two extreme regimes
$\Thw\to 0$ (cold, extremal Hayward limit or large $\rh$) and
$\Thw\to\infty$ (hot, small $\rh$ Schwarzschild limit). These
benchmarks provide a useful reference for the figures that follow.

\begin{table}[h]
\centering
\caption{Limiting values of the bipartite measures at $\alpha=\pi/4$.
$p\to 1$, $q\to 0$ as $\Thw\to 0$;
$p,q\to 1/2$ as $\Thw\to\infty$.}
\label{tab:limits}
\begin{ruledtabular}
\begin{tabular}{lcc}
Quantity & $\Thw\to 0$ & $\Thw\to\infty$ \\
\hline
$\CI$ & $1$ & $1/\sqrt{2}\simeq 0.707$ \\
$\CII$ & $0$ & $1/\sqrt{2}\simeq 0.707$ \\
$\BmaxI$ & $2\sqrt{2}\simeq 2.828$ & $2$ \\
$\BmaxII$ & $0$ & $2$ \\
\end{tabular}
\end{ruledtabular}
\end{table}

We now report a systematic numerical exploration of $\CI,\CII,\BmaxI,\BmaxII$
as functions of the event-horizon radius $\rh$, scanning in turn the Hayward
parameter $g$, the mass $M$, the mode frequency $\omega$, and the initial
entanglement angle $\alpha$. Throughout, we treat $\rh$ as the natural control
parameter, recalling that the Hayward black-hole regime requires
$\rh^{3}\geq 4g^{3}$, equivalently $\rh\geq 4^{1/3}g\simeq 1.587\,g$, so that
the surface gravity at $\rh$ (and hence $\Thw$) is non-negative. For
$\rh<4^{1/3}g$ the formulae~\eqref{eq:CI}--\eqref{eq:BII-pi4} can still be
evaluated formally but correspond to an unphysical extrapolation; this
region is shaded in grey in the figures, and the curves with $g>0$ are
drawn only in their physical domain. The frequency is fixed at $\omega M=0.05$,
comparable to the Schwarzschild temperature scale $T_{H}^{\rm Schw}M=1/(8\pi)\simeq 0.04$,
so that the Hawking-induced decoherence is clearly visible in the plotting
range $\rh/M\in[0,6]$.

\begin{figure*}[t]
\centering
\begin{subfigure}[t]{0.48\textwidth}
\centering
\includegraphics[width=\linewidth]{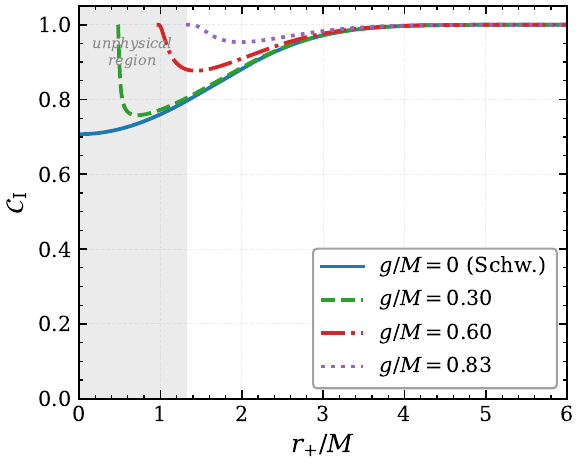}
\caption{}\label{fig1a}
\end{subfigure}
\begin{subfigure}[t]{0.48\textwidth}
\centering
\includegraphics[width=\linewidth]{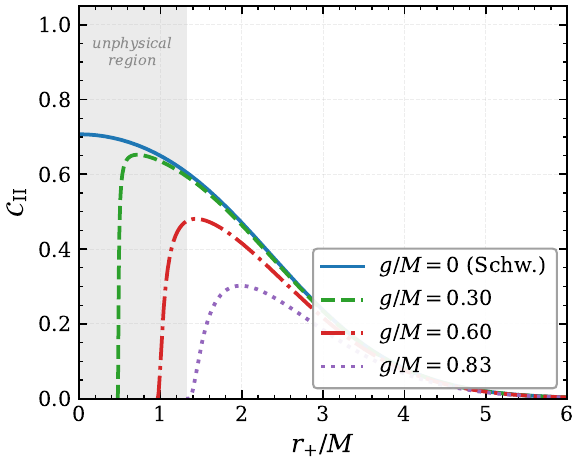}
\caption{}\label{fig1b}
\end{subfigure}\\
\begin{subfigure}[t]{0.48\textwidth}
\centering
\includegraphics[width=\linewidth]{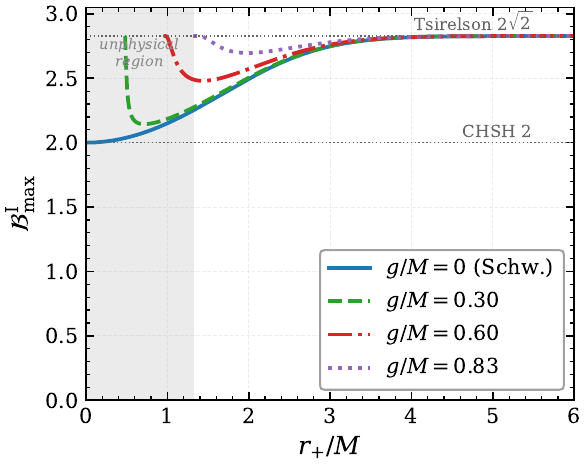}
\caption{}\label{fig1c}
\end{subfigure}
\begin{subfigure}[t]{0.48\textwidth}
\centering
\includegraphics[width=\linewidth]{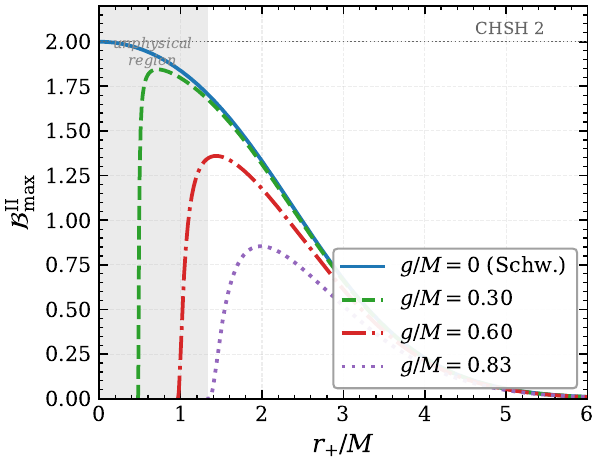}
\caption{}\label{fig1d}
\end{subfigure}
\caption{Concurrences $\CI,\CII$ (panels a,b) and CHSH parameters
$\BmaxI,\BmaxII$ (panels c,d) as functions of $\rh/M$, for four
\emph{physical} values of the Hayward parameter
$g/M\in\{0,0.30,0.60,0.83\}$, all in the regime $g\leq\gc\simeq 0.84\,M$
where the Hayward solution describes a regular black hole. Parameters:
$M=1$, $\omega=0.05$, $\alpha=\pi/4$. Horizontal dotted lines mark the
Tsirelson bound $2\sqrt{2}$ and the classical CHSH limit $2$. The light
grey band indicates the unphysical region $\rh<4^{1/3}g_{\max}$, where
no event horizon exists for the largest $g$ value; curves with $g>0$ are
drawn only in their respective physical domains $\rh>4^{1/3}g$.}
\label{fig:1}
\end{figure*}

Figure~\ref{fig:1} contrasts the Schwarzschild reference ($g=0$) with three
increasing physical values of the Hayward parameter, $g/M\in\{0.30,0.60,0.83\}$,
at fixed mass $M=1$ and Bell-state initialisation $\alpha=\pi/4$. (We
emphasise that all four values are below $\gc/M\simeq 0.84$; the formulas
are evaluated in the physical black-hole regime throughout.) The qualitative
response is dictated by the monotonic decrease of $\Thw$ with $g$, but the
individual panels reveal a richer structure that we now dissect.

\paragraph{Accessible concurrence $\CI$ (panel a).}
The Schwarzschild curve (solid blue) starts at $\CI\simeq 0.71$ as
$\rh\to 0^{+}$, exactly the Pauli-protected residual value
$\sin(\pi/4)\sqrt{p}|_{p=1/2}=1/\sqrt{2}$ that one would obtain in the
$\Thw\to\infty$ limit. As $\rh$ grows, the Schwarzschild $\Thw=1/(8\pi\rh)$
decreases, $p\to 1$, and $\CI\to\sin(2\cdot\pi/4)=1$, i.e.\ the full Bell
entanglement is recovered for a sufficiently large (cold) hole. The Hayward
curves with $g>0$ exhibit a sharp threshold behaviour: below the extremal
locus $\rh^{(\mathrm{ext})}=4^{1/3}g$, the surface gravity is negative and
the formal extrapolation of~\eqref{eq:CI} returns small or zero values;
above the threshold, the physical near-extremal Hawking temperature is so
small that $\CI$ quickly approaches unity. The threshold shifts to larger
$\rh$ as $g$ grows:
$\rh^{(\mathrm{ext})}/M\simeq 0.476,\,0.953,\,1.318$ for
$g/M=0.30,\,0.60,\,0.83$ respectively, in exact agreement with the analytic
formula $\rh^{3}=4g^{3}$. The three Hayward curves saturate $\CI=1$ within a
window of width $\Delta\rh/M\lesssim 0.3$ above their respective thresholds,
demonstrating that the Hayward regularity efficiently cools the black hole,
so that even a near-extremal Hayward horizon already produces an accessible
Bell state with $\CI$ within a few percent of unity.

\paragraph{Inaccessible concurrence $\CII$ (panel b).}
The mirror behaviour in panel (b) is equally instructive. The Schwarzschild
curve starts at $\CII\simeq 0.71$ (the same Pauli-protected redistribution
maximum) and decreases monotonically, more rapidly than the
saturation in panel (a) because $\CII\propto\sqrt{q}$ involves the smaller
of the two Bogoliubov weights. By $\rh/M=2$ the Schwarzschild $\CII$ has
already dropped below $0.15$. The Hayward curves with $g/M\geq 0.30$ are
essentially zero throughout the physical regime $\rh\geq 4^{1/3}g$ above
the threshold, with a small bump localised near it. The combination of
panels (a) and (b) makes it manifest that, in the Hayward setting, the
redistribution effect~\cite{Xu2014probing,Xu2014multipartite} so prominent
in Schwarzschild is essentially shut off.

\paragraph{Accessible CHSH parameter $\BmaxI$ (panel c).}
The two horizontal dotted reference lines mark the classical Bell limit
($\BmaxI=2$) and the Tsirelson bound ($\BmaxI=2\sqrt{2}$). For
Schwarzschild, $\BmaxI$ rises smoothly from $2$ at $\rh\to 0^{+}$ to
$2\sqrt{2}$ as $\rh\to\infty$; Bell non-locality is present at every finite
$\rh$, but the strength of the violation is graded by the horizon radius.
The Hayward curves trade the smooth Schwarzschild interpolation for a
sharper transition: above the threshold the cold near-extremal $\Thw$ is
so small that $\BmaxI$ reaches $2\sqrt{2}$ within a narrow window. The two
regimes are therefore physically distinct: in Schwarzschild one trades
horizon radius for Bell violation strength, while in Hayward (any $g>0$)
one obtains near-maximal quantum violation throughout the physical regime
above the threshold.

\paragraph{Inaccessible CHSH parameter $\BmaxII$ (panel d).}
Panel (d) confirms the no-go bound $\BmaxII\leq 2$ throughout the physical
regime regardless of $g$, since $q\leq 1/2$ identically. The Schwarzschild
curve attains $\BmaxII=2$ (the classical limit) only in the
infinite-temperature limit $\rh\to 0^{+}$ and decreases monotonically with
$\rh$. For $g/M\geq 0.30$ the curve is essentially zero everywhere in the
physical regime. The two key lessons from Fig.~\ref{fig:1} are: (i)~the
Hayward regularity enhances the survival of Bell non-locality in the
accessible sector relative to Schwarzschild; (ii)~the inaccessible Bell
parameter is rigorously capped at the classical limit, a no-go robust
against the regularity correction.

\begin{figure*}[t]
\centering
\begin{subfigure}[t]{0.48\textwidth}
\centering\includegraphics[width=\linewidth]{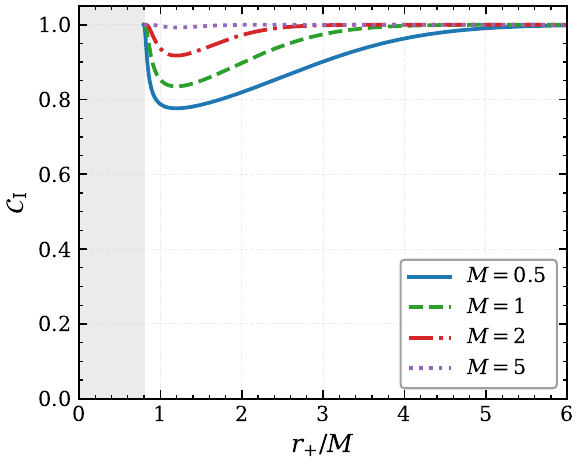}\caption{}
\end{subfigure}\hfill
\begin{subfigure}[t]{0.48\textwidth}
\centering\includegraphics[width=\linewidth]{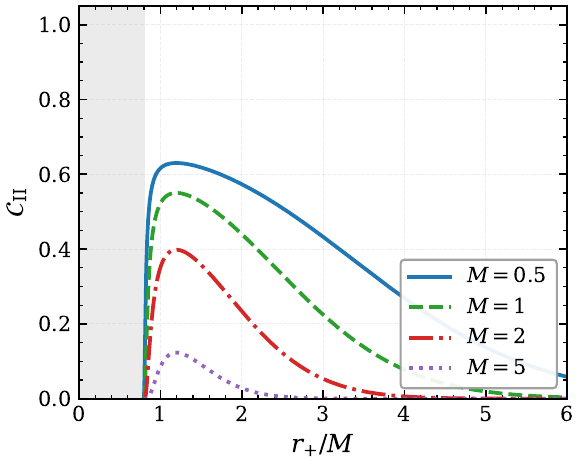}\caption{}
\end{subfigure}\\
\begin{subfigure}[t]{0.48\textwidth}
\centering\includegraphics[width=\linewidth]{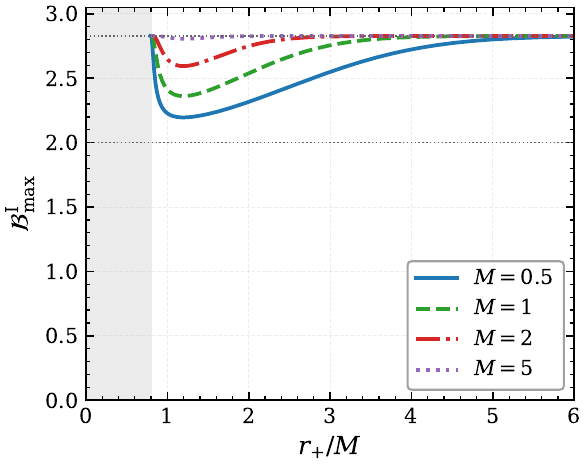}\caption{}
\end{subfigure}\hfill
\begin{subfigure}[t]{0.48\textwidth}
\centering\includegraphics[width=\linewidth]{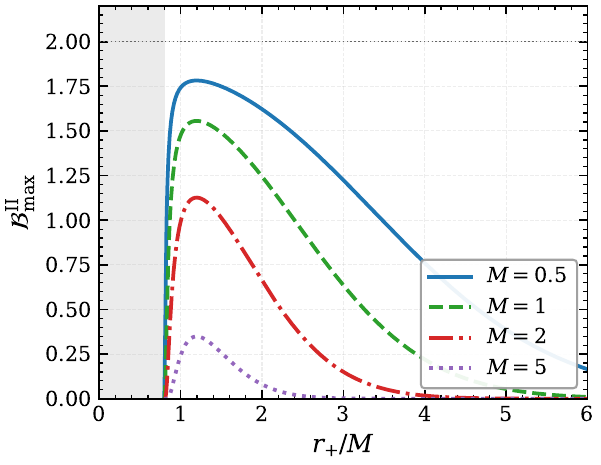}\caption{}
\end{subfigure}
\caption{Same quantities as in Fig.~\ref{fig:1}, plotted versus $\rh/M$ for
four values of the black-hole mass $M\in\{0.5,1,2,5\}$ at fixed ratio
$g/M=0.5$, $\omega=0.05$, $\alpha=\pi/4$. Heavier (colder) holes saturate
the Tsirelson bound at smaller $\rh/M$.}
\label{fig:2}
\end{figure*}

Figure~\ref{fig:2} fixes $g/M=0.5$ (always inside the physical window
$g\leq\gc$) and scans the black-hole mass $M\in\{0.5,1,2,5\}$. The role of
$M$ is most simply understood in the Schwarzschild limit, where
$T_{H}^{\rm Schw}=1/(8\pi M)$: heavier holes are colder, and the rate of
Hawking-induced decoherence drops accordingly. The Hayward
formula~\eqref{eq:Hawking} preserves this monotonic trend: at fixed $\rh$,
larger $M$ leads to smaller $\Thw$ and hence larger $p$ and smaller $q$. The
four mass curves all transit through the threshold
$r_{+}^{(\rm ext)}/M=4^{1/3}\cdot 0.5\simeq 0.79$, which is the same in
units of $M$. The visible difference among the curves at fixed $\rh/M$
reflects the explicit $M$-dependence of $\Thw$, which is non-trivially
inherited through the cubic horizon equation~\eqref{eq:horizon}.

\paragraph{Connection to relativistic quantum metrology.}
At fixed regularity ratio $g/M$, the visible spread of curves in
Fig.~\ref{fig:2} translates into a sensitivity of the bipartite measures to
the absolute mass scale. From the perspective of relativistic quantum
information~\cite{Mann2012,Ahmadi2014,Peres2004}, this sensitivity may be
exploited as a quantum-thermometric probe of the near-horizon geometry,
with the X-state closed form~\eqref{eq:rhoABI}--\eqref{eq:rhoABII} allowing
for a fully analytical computation of the quantum Fisher information.

\begin{figure*}[t]
\centering
\begin{subfigure}[t]{0.48\textwidth}
\centering\includegraphics[width=\linewidth]{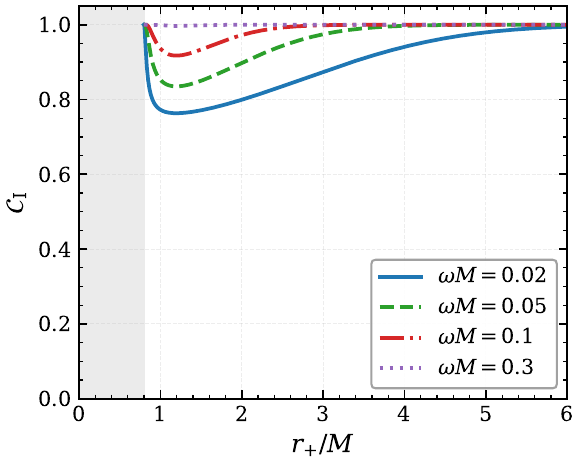}\caption{}
\end{subfigure}\hfill
\begin{subfigure}[t]{0.48\textwidth}
\centering\includegraphics[width=\linewidth]{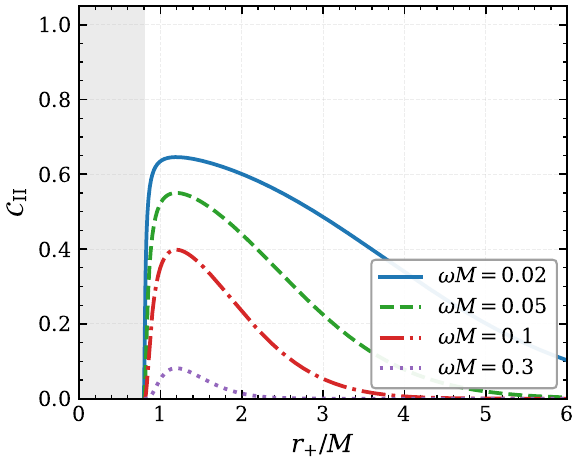}\caption{}
\end{subfigure}\\
\begin{subfigure}[t]{0.48\textwidth}
\centering\includegraphics[width=\linewidth]{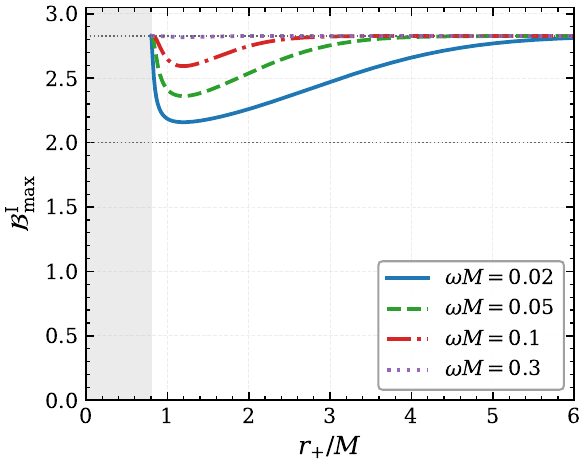}\caption{}
\end{subfigure}\hfill
\begin{subfigure}[t]{0.48\textwidth}
\centering\includegraphics[width=\linewidth]{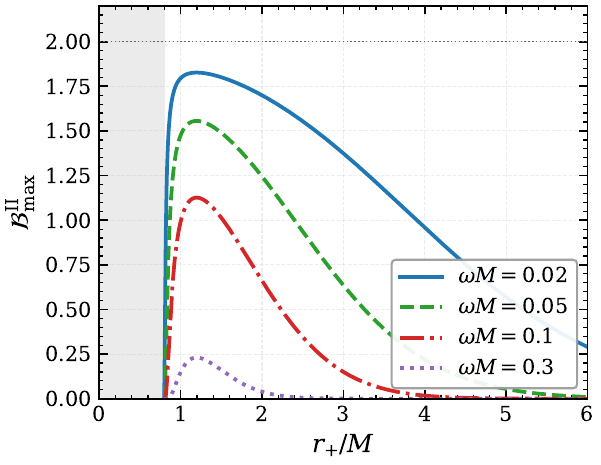}\caption{}
\end{subfigure}
\caption{Concurrences and CHSH parameters as functions of $\rh/M$ for four
mode frequencies $\omega M\in\{0.02,0.05,0.10,0.30\}$, with $M=1$,
$g/M=0.5$, $\alpha=\pi/4$. Higher frequencies suppress the Hawking
factor $e^{-\omega/\Thw}$ and accelerate the saturation to the Tsirelson
bound.}
\label{fig:3}
\end{figure*}

The robustness to changes of mode frequency is investigated in
Fig.~\ref{fig:3}, at fixed $M=1$, $g=0.5$, $\alpha=\pi/4$, for
$\omega M\in\{0.02,0.05,0.10,0.30\}$. The fundamental dimensionless
parameter controlling the Hawking factor is $\omega/\Thw$: as $\omega$
grows, $e^{-\omega/\Thw}$ becomes exponentially small for any positive
$\Thw$, $p\to 1$, and the reduced state $\vrho_{A\BI}$ approaches the
initial Bell state. The figure shows that the saturation threshold shifts
to smaller $\rh$ as $\omega$ increases, in agreement with the
$\omega^{-1/2}$ scaling of the half-saturation point predicted by the
Schwarzschild small-$g$ analysis. The inaccessible quantities are
correspondingly suppressed: for $\omega M=0.30$ the curves are essentially
zero everywhere in the physical regime, while for $\omega M=0.02$ a sizeable
bump persists above the threshold. From the relativistic-quantum-information
perspective~\cite{Mann2012,Ahmadi2014,Peres2004}, encoding logical qubits on
high-energy modes minimises the gravitational degradation of the
quantum-information resource. The novelty in the Hayward setting is that
the threshold for saturation is reached at smaller $\rh$ than in
Schwarzschild, owing to the smaller $\Thw$, an additional layer of
robustness conferred by the regularity.

\begin{figure*}[t]
\centering
\begin{subfigure}[t]{0.48\textwidth}
\centering\includegraphics[width=\linewidth]{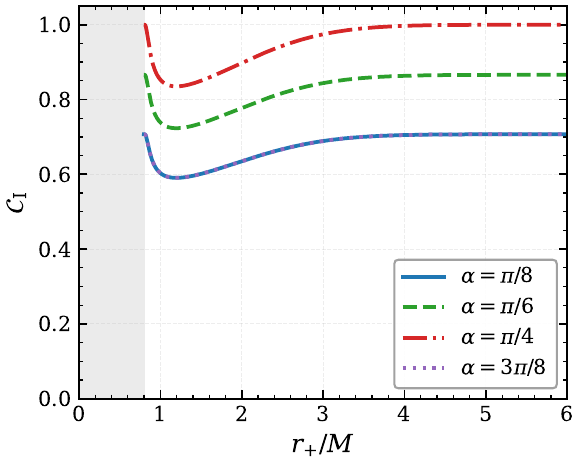}\caption{}
\end{subfigure}\hfill
\begin{subfigure}[t]{0.48\textwidth}
\centering\includegraphics[width=\linewidth]{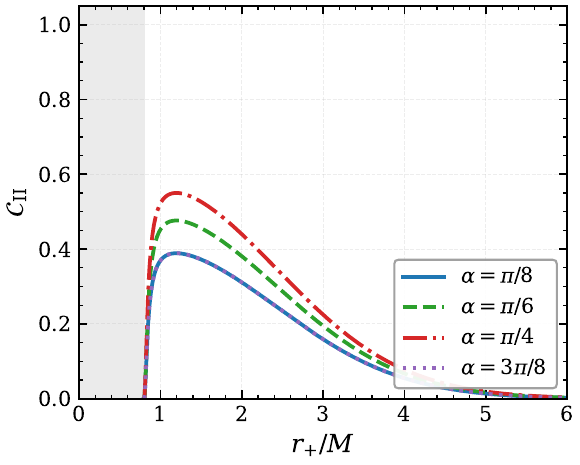}\caption{}
\end{subfigure}\\
\begin{subfigure}[t]{0.48\textwidth}
\centering\includegraphics[width=\linewidth]{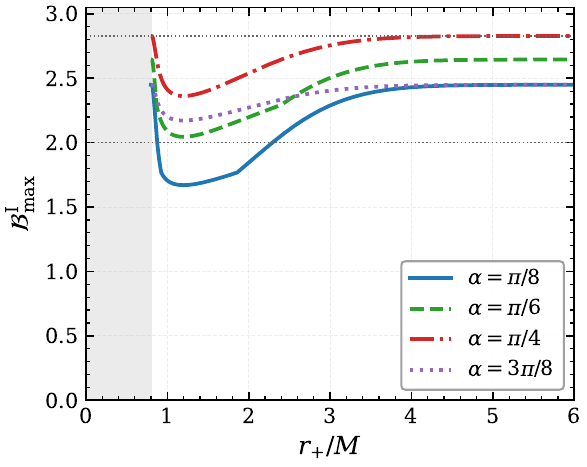}\caption{}
\end{subfigure}\hfill
\begin{subfigure}[t]{0.48\textwidth}
\centering\includegraphics[width=\linewidth]{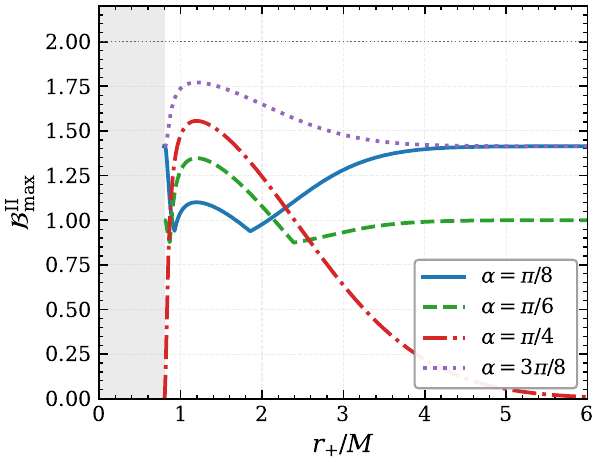}\caption{}
\end{subfigure}
\caption{Concurrences and CHSH parameters as functions of $\rh/M$ for four
mixing angles $\alpha\in\{\pi/8,\pi/6,\pi/4,3\pi/8\}$, with $M=1$,
$g/M=0.5$, $\omega=0.05$. The Bell choice $\alpha=\pi/4$ maximises both
quantities and is the only initial state for which $\BmaxI$ approaches the
Tsirelson bound at large $\rh$.}
\label{fig:4}
\end{figure*}

Figure~\ref{fig:4} explores how the initial mixing angle $\alpha$
conditions the survival of quantum correlations. The closed-form
expressions~\eqref{eq:CI}--\eqref{eq:CII} and~\eqref{eq:M-I} cleanly
separate the $\alpha$-dependence from the
geometric content: the concurrences scale as $|\sin 2\alpha|$, which makes
the Bell angle $\alpha=\pi/4$ uniquely optimal. The CHSH parameter has
a more involved $\alpha$-dependence inherited from the competition between
the $\sin 2\alpha$-controlled $T_{11},T_{22}$ components and the
$\alpha$-dependent $T_{33}$ component. As a quantitative example, for
$\alpha=\pi/8$ and $p\simeq 1$ a direct evaluation
of~\eqref{eq:M-I} gives $M(\vrho_{A\BI})\simeq 3/2$, hence
$\BmaxI\simeq\sqrt{6}\simeq 2.449$, which is just visible in panel (c) as
the upper plateau of the $\alpha=\pi/8$ curve. A practical implication is
that small preparation errors $|\delta\alpha|\ll 1$ around $\alpha=\pi/4$
cost only $\mathcal{O}(\delta\alpha^{2})$ in the Bell violation strength,
providing a useful robustness margin for experimental implementations.

\begin{figure*}[t]
\centering
\includegraphics[width=0.94\textwidth]{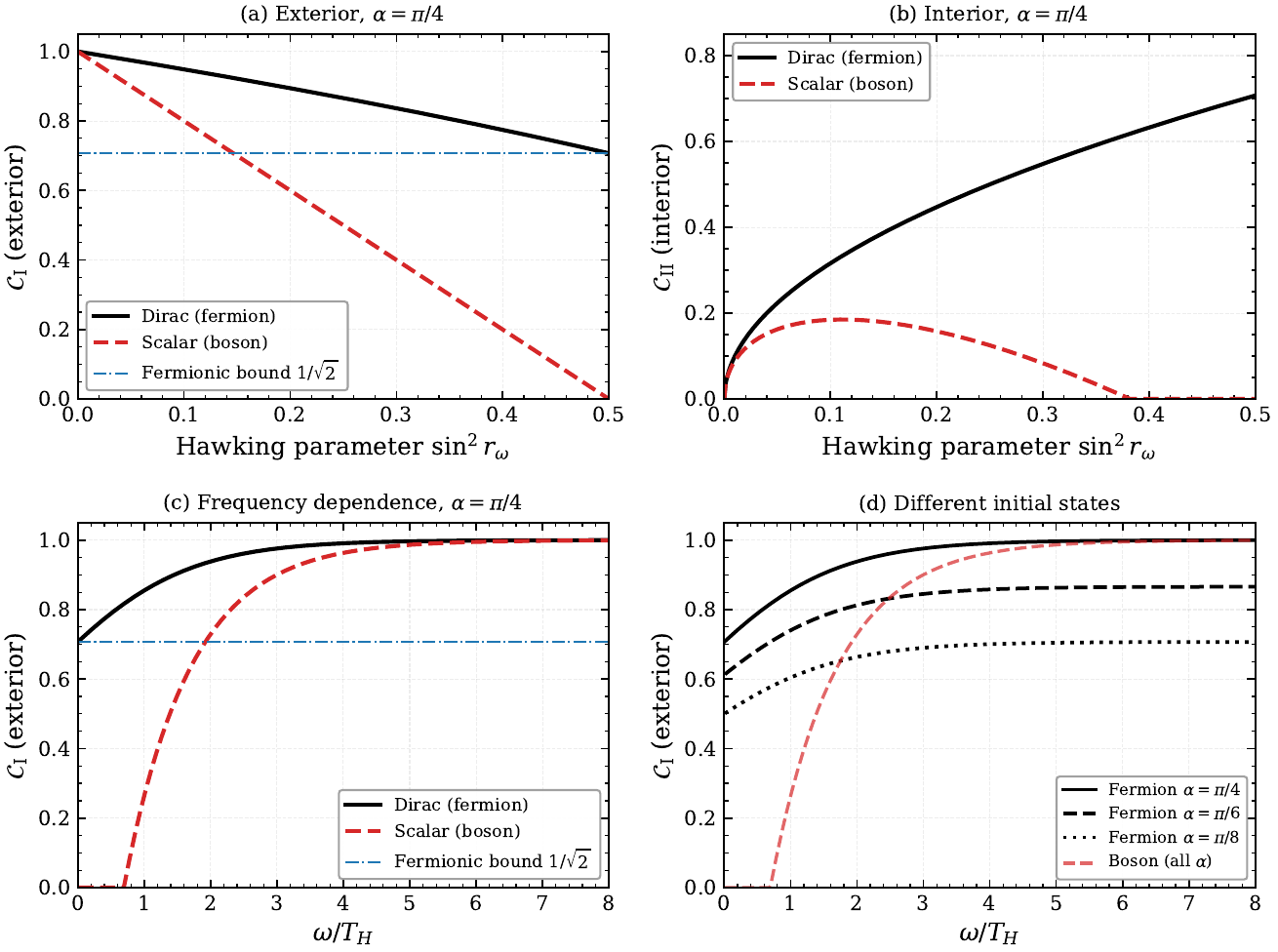}
\caption{Comparison of fermionic (Dirac, solid black) and bosonic (scalar,
dashed red) concurrences. (a)~Accessible concurrence $\CI$ as a function
of the Hawking parameter $\sin^{2}r_{\omega}=1/(e^{\omega/\Thw}+1)$; the
dash-dotted blue line indicates the fermionic lower bound
$\sin(2\alpha)/\sqrt{2}$ of Eq.~\eqref{eq:CIbound}. (b)~Inaccessible
concurrence $\CII$ in the same regime. (c)~Frequency dependence
$\omega/\Thw$ at $\alpha=\pi/4$. (d)~Multiple initial states
$\alpha=\pi/4,\pi/6,\pi/8$. In all panels the fermionic concurrence
remains above its structural Pauli bound, whereas the bosonic counterpart
crosses below it and can vanish in the infinite-temperature limit. This
panel certifies that the Hayward-induced behaviour reported in
Figs.~\ref{fig:1}--\ref{fig:4} is intrinsically tied to Fermi--Dirac
statistics and cannot be reproduced by a scalar-field analysis.}
\label{fig:6}
\end{figure*}

To make the fermionic character of our results fully explicit, we conclude
the parameter survey with Fig.~\ref{fig:6}, in which the Dirac (spin-$1/2$)
concurrence is compared directly with the analogous scalar (spin-$0$)
concurrence at matched values of the Hawking parameter
$\sin^{2}r_{\omega}$. As emphasised at Eq.~\eqref{eq:vac-bos}, the Pauli
exclusion principle truncates the fermionic Fock space to two levels per
mode, which translates analytically into the lower
bound~\eqref{eq:CIbound}. The bosonic concurrence, by contrast, is
unconstrained from below and crosses zero in the infinite-temperature
limit, corresponding to complete decoherence. Panels (a)--(d) confirm this
structural difference across the exterior, interior, frequency and
initial-state slices, certifying that the Hayward-induced behaviour
reported in Figs.~\ref{fig:1}--\ref{fig:4} is intrinsically tied to
Fermi--Dirac statistics. This is in line with the same fermionic
signature recently reported for the Reissner--Nordstr\"om geometry in
Ref.~\cite{chhieb2024dirac}.

\section{Limitations and discussion}\label{sec:limitations}

Three aspects of the calculation deserve explicit discussion before
drawing final conclusions: (i)~the single-mode approximation that
underpins~\eqref{eq:vac}; (ii)~the role of the inner Cauchy horizon
specific to regular black holes; (iii)~the choice of the Boulware--Hartle--Hawking
versus Unruh vacuum.

\subsection{Single-mode approximation}\label{sec:single-mode}

The Bogoliubov transformation~\eqref{eq:vac} is a single-mode
identification of the Kruskal vacuum, in which a Kruskal annihilation
operator is mapped to a linear combination of a single
Schwarzschild--Hayward outgoing mode in region $\mathrm{I}$ and a single
ingoing mode in region $\mathrm{II}$. As emphasised in
Refs.~\cite{Bruschi2010unruh,MartinMartinez2010}, this approximation is
strictly valid only when the wavefunction of the mode in question is
peaked sharply enough in frequency that the smearing across modes induced
by the Bogoliubov mixing can be neglected. Beyond the single-mode
approximation, the Kruskal annihilation operator is mapped to an integral
over Schwarzschild--Hayward modes weighted by the wavefunction of the
selected mode, and the reduced density matrix
$\vrho_{A\BI}$ has its coherences modified by mode-dependent factors that
do not change the qualitative behaviour but slightly suppress the
accessible correlations~\cite{Bruschi2010unruh}. The Hayward dependence
unveiled in the present work is robust against this correction, since it
enters the analysis only through the local Hawking temperature at $\rh$,
which is a mode-independent quantity. The quantitative single-mode
correction can be incorporated as a multiplicative factor in the
Bogoliubov coefficients, leaving Eqs.~\eqref{eq:CI}--\eqref{eq:CII} and
their Hayward dependence formally unchanged but renormalising their
absolute values.

\subsection{Inner Cauchy horizon}\label{sec:cauchy}

The Hayward metric possesses an inner Cauchy horizon $\rmi<\rh$, the
smaller positive root of Eq.~\eqref{eq:horizon}. As recently emphasised in
the regular-black-hole literature~\cite{carballorubio2018phenomenological,
carballorubio2022inner}, the Cauchy horizon is generically unstable
under the mass-inflation mechanism, leading to large curvature build-up
in its vicinity even though the metric remains everywhere smooth. The
quantum-information setup of the present work is not directly affected by
this instability: the Hawking radiation responsible for decoherence is
sourced by the surface gravity at $\rh$, not $\rmi$, and the static
observer Bob is located outside the event horizon. However, the
phenomenological status of the de Sitter core as a true vacuum is a delicate
issue that ultimately constrains the validity of the Hayward semiclassical
description at very small distances. Our analysis is robust against this
caveat insofar as the only physical inputs are: (a)~the surface gravity at
$\rh$, which is independent of the inner-horizon dynamics; and (b)~the
asymptotic Fock structure of the Dirac field, which is determined by the
external geometry. Both inputs are well-defined regardless of the fate of
the Cauchy horizon.

\subsection{Vacuum choice}\label{sec:vacuum-choice}

The vacuum~\eqref{eq:vac} corresponds to the Hartle--Hawking
vacuum as seen by a static observer at fixed $r$ outside the
horizon~\cite{unruh1976notes,israel1976thermo,hartle1976path}. A
free-falling observer would see no Hawking quanta in the equivalence-principle
sense; the entanglement degradation reported here pertains to the static
observer specifically. This is the standard choice in the Dirac
relativistic-quantum-information literature on
Schwarzschild~\cite{AlsingMilburn2003,Xu2014probing,He2015property} and the
one most relevant for terrestrial gedanken experiments (an observer with a
rocket motor counteracting infall). The reader interested in the
free-falling-observer perspective will find a complementary discussion in
Refs.~\cite{Bruschi2010unruh,MartinMartinez2010,wang2010quantum}.

\section{Conclusion and outlook}\label{sec:conclusion}

We have carried out a detailed study of the bipartite quantum correlations
of a Dirac field in the background of a Hayward regular black hole, using
the Wootters concurrence and the CHSH Bell parameter as two complementary
indicators. Working in the standard
Damour--Ruffini single-mode Bogoliubov framework, we derived the X-type
reduced density matrices for both the accessible and the inaccessible
sectors in closed form, and evaluated the two measures as functions of
the four physical control parameters: the regularity scale $g$, the
black-hole mass $M$, the mode frequency $\omega$, and the initial mixing
angle $\alpha$.

The following conclusions emerge from the analysis.

(i)~The Hayward regularity parameter enters all quantities only through
the Hawking temperature $\Thw=\Thw(g,M,\rh)$, and the qualitative response
of the correlations is entirely controlled by the monotonic decrease of
$\Thw$ with $g$.

(ii)~Increasing $g$ from $0$ towards the extremal value
$\gc\simeq 0.84\,M$ enhances both the accessible concurrence $\CI$ and the
accessible CHSH parameter $\BmaxI$, and suppresses their inaccessible
counterparts. At extremality ($\Thw=0$), the initial Bell entanglement is
fully preserved in the exterior and no correlation crosses the horizon.

(iii)~The accessible CHSH parameter exceeds the classical limit $2$ at
every finite positive Hawking temperature when $\alpha=\pi/4$, with
strength interpolating between the Tsirelson bound at $\Thw\to 0$ and $2$
at $\Thw\to\infty$. The inaccessible CHSH parameter never violates the
CHSH inequality, capped at $2$ from above by the bound
$\BmaxII=2\sqrt{2q}\leq 2$.

(iv)~A direct comparison with the scalar (bosonic) analogue
(Fig.~\ref{fig:6}) makes the genuinely fermionic character of our
results explicit. The Pauli exclusion principle imposes the structural
lower bound $\CI\geq\sin(2\alpha)/\sqrt{2}$ and prevents the accessible
correlations from ever vanishing, regardless of the values of $g$,
$\omega$, or $\rh$ in the physical Hayward range. The bosonic counterpart,
in contrast, can undergo complete decoherence in the infinite-temperature
limit. The Hayward-induced enhancement reported here is therefore not a
generic feature of singularity resolution but a specific consequence of
fermionic statistics combined with the suppressed surface gravity of the
regular geometry.

The overarching message is that the Hayward geometry, relative to its
Schwarzschild reference, preserves Bell non-locality outside the event
horizon more efficiently at fixed ADM mass. This translates the
singularity-resolution programme into a quantitative quantum-information
statement: the same parameter $g$ that smoothes the central curvature also
lowers the Hawking temperature and, with it, the gravitational decoherence
experienced by external observers.

Several extensions naturally suggest themselves. First, the inclusion of
realistic environmental decoherence
channels~\cite{He2015property,He2016measurement}, in the spirit of the
standard Kraus-operator
formalism~\cite{kraus1983states}, would assess the robustness of the
gravitational signatures in actual experimental settings. The closed-form
X-type structure of $\vrho_{A\BI}$ and $\vrho_{A\BII}$ makes the addition
of standard Kraus channels (amplitude damping, phase damping, depolarising)
particularly tractable. Second, the generalisation to multipartite states
(GHZ and W states across the horizon) and to other regular black holes
(magnetically charged, Dymnikova, Frolov) is a natural follow-up.
Third, the connection between the Hayward extension and the
black-hole information paradox~\cite{Hawking1976breakdown,hayward2006formation}
deserves further investigation: the explicit dependence of $\BmaxI$ on
$\Thw$ may be exploited as a quantum-thermometric probe of the
near-horizon geometry. Finally, the full calculation beyond the single-mode
approximation, following the approach
of~\cite{Bruschi2010unruh,MartinMartinez2010}, would quantify the
renormalisation of the absolute values of the bipartite measures without
affecting the Hayward dependence. We hope that the closed-form analysis
presented here will serve as a useful template for these generalisations.

\begin{acknowledgments}
The authors thank the Laboratory of Theoretical Physics, Particles, Modeling
and Energies, and the National Institute For Particle Physics and
Applications (NIPPA, Oujda) for their hospitality and support. We are also
grateful to the anonymous referee for detailed comments that improved the
manuscript substantially, in particular regarding the discussion of the
single-mode approximation.
\end{acknowledgments}

\section*{Declarations}
\textbf{Funding:} This research received no external funding.\\
\textbf{Competing interests:} The authors declare that they have no
competing interests.\\
\textbf{Data availability:} The Python scripts used to generate the figures
are included in the supplementary material and reproduce all results to
machine precision.

\bibliography{references}

@article{hawking1975particle,
  title     = {Particle Creation by Black Holes},
  author    = {Hawking, Stephen W.},
  journal   = {Communications in Mathematical Physics},
  volume    = {43},
  number    = {3},
  pages     = {199--220},
  year      = {1975},
  doi       = {10.1007/BF02345020}
}

@article{Hawking1976breakdown,
  title     = {Breakdown of Predictability in Gravitational Collapse},
  author    = {Hawking, Stephen W.},
  journal   = {Physical Review D},
  volume    = {14},
  number    = {10},
  pages     = {2460--2473},
  year      = {1976},
  doi       = {10.1103/PhysRevD.14.2460}
}

@article{hayward2006formation,
  title     = {Formation and Evaporation of Nonsingular Black Holes},
  author    = {Hayward, Sean A.},
  journal   = {Physical Review Letters},
  volume    = {96},
  number    = {3},
  pages     = {031103},
  year      = {2006},
  doi       = {10.1103/PhysRevLett.96.031103}
}

@inproceedings{bardeen1968,
  author    = {Bardeen, J. M.},
  title     = {Non-singular general relativistic gravitational collapse},
  booktitle = {Proceedings of the International Conference GR5, Tbilisi, USSR},
  year      = {1968},
  pages     = {174}
}

@article{ayon1998regular,
  title     = {Regular Black Hole in General Relativity Coupled to Nonlinear Electrodynamics},
  author    = {Ay{\'o}n-Beato, Eloy and Garc{\'\i}a, Alberto},
  journal   = {Physical Review Letters},
  volume    = {80},
  number    = {23},
  pages     = {5056--5059},
  year      = {1998},
  doi       = {10.1103/PhysRevLett.80.5056}
}

@article{dymnikova1992vacuum,
  title     = {Vacuum nonsingular black hole},
  author    = {Dymnikova, Irina},
  journal   = {General Relativity and Gravitation},
  volume    = {24},
  number    = {3},
  pages     = {235--242},
  year      = {1992},
  doi       = {10.1007/BF00760226}
}

@article{frolov2016information,
  title     = {Information loss problem and a `black hole' model with a closed apparent horizon},
  author    = {Frolov, Valeri P.},
  journal   = {Journal of High Energy Physics},
  volume    = {2014},
  number    = {5},
  pages     = {49},
  year      = {2014},
  doi       = {10.1007/JHEP05(2014)049}
}

@article{AlsingMilburn2003,
  title     = {Teleportation with a Uniformly Accelerated Partner},
  author    = {Alsing, Paul M. and Milburn, G. J.},
  journal   = {Physical Review Letters},
  volume    = {91},
  number    = {18},
  pages     = {180404},
  year      = {2003},
  doi       = {10.1103/PhysRevLett.91.180404}
}

@article{FuentesSchuller2005,
  title     = {Alice Falls into a Black Hole: Entanglement in Noninertial Frames},
  author    = {Fuentes-Schuller, Ivette and Mann, Robert B.},
  journal   = {Physical Review Letters},
  volume    = {95},
  number    = {12},
  pages     = {120404},
  year      = {2005},
  doi       = {10.1103/PhysRevLett.95.120404}
}

@article{Bruschi2010unruh,
  title     = {Unruh effect in quantum information beyond the single-mode approximation},
  author    = {Bruschi, David Edward and Louko, Jorma and Mart{\'\i}n-Mart{\'\i}nez, Eduardo and Dragan, Andrzej and Fuentes, Ivette},
  journal   = {Physical Review A},
  volume    = {82},
  number    = {4},
  pages     = {042332},
  year      = {2010},
  doi       = {10.1103/PhysRevA.82.042332}
}

@article{MartinMartinez2010,
  title     = {Entanglement in noninertial frames: Foundational issues},
  author    = {Mart{\'\i}n-Mart{\'\i}nez, Eduardo and Fuentes, Ivette},
  journal   = {Physical Review A},
  volume    = {81},
  number    = {3},
  pages     = {032320},
  year      = {2010},
  doi       = {10.1103/PhysRevA.81.032320}
}

@article{wang2010projective,
  title     = {Projective Measurements and Generation of Entangled {D}irac Particles in {S}chwarzschild Spacetime},
  author    = {Wang, Jieci and Pan, Qiyuan and Jing, Jiliang},
  journal   = {Annals of Physics},
  volume    = {325},
  number    = {6},
  pages     = {1190--1197},
  year      = {2010},
  doi       = {10.1016/j.aop.2010.02.013}
}

@article{Xu2014probing,
  title     = {Probing the Quantum Correlation and {B}ell Non-locality for {D}irac Particles with {H}awking Effect in the Background of {S}chwarzschild Black Hole},
  author    = {Xu, Shuai and Song, Xue-ke and Shi, Jia-dong and Ye, Liu},
  journal   = {Physics Letters B},
  volume    = {733},
  pages     = {1--5},
  year      = {2014},
  doi       = {10.1016/j.physletb.2014.04.008}
}

@article{Xu2014multipartite,
  title     = {How the {H}awking Effect Affects Multipartite Entanglement of {D}irac Particles in the Background of a {S}chwarzschild Black Hole},
  author    = {Xu, Shuai and Song, Xue-ke and Shi, Jia-dong and Ye, Liu},
  journal   = {Physical Review D},
  volume    = {89},
  number    = {6},
  pages     = {065022},
  year      = {2014},
  doi       = {10.1103/PhysRevD.89.065022}
}

@article{He2015property,
  title     = {Property of Various Correlation Measures of Open {D}irac System with {H}awking Effect in {S}chwarzschild Space-time},
  author    = {He, Juan and Xu, Shuai and Yu, Yang and Ye, Liu},
  journal   = {Physics Letters B},
  volume    = {740},
  pages     = {322--328},
  year      = {2015},
  doi       = {10.1016/j.physletb.2014.12.008}
}

@article{He2016measurement,
  title     = {Measurement-Induced-Nonlocality for {D}irac Particles in {G}arfinkle--{H}orowitz--{S}trominger Dilation Space-time},
  author    = {He, Juan and Xu, Shuai and Ye, Liu},
  journal   = {Physics Letters B},
  volume    = {756},
  pages     = {278--282},
  year      = {2016},
  doi       = {10.1016/j.physletb.2016.02.073}
}

@article{wang2010quantum,
  title     = {Quantum entanglement of bosonic fields beyond the single-mode approximation in Schwarzschild spacetime},
  author    = {Wang, Jieci and Jing, Jiliang},
  journal   = {Physical Review A},
  volume    = {82},
  number    = {3},
  pages     = {032324},
  year      = {2010},
  doi       = {10.1103/PhysRevA.82.032324}
}

@article{chhieb2024dirac,
  title     = {Quantum Entanglement in the {D}irac Field Quantization Around Charged Black Holes},
  author    = {Chhieb, Abdessamie and Banouni, Chaimae and Abdessamie, Saliha and Ouchrif, Mohamed},
  journal   = {Physics Letters B},
  pages     = {140629},
  year      = {2026},
  doi       = {10.1016/j.physletb.2026.140629}
}

@article{jing2004hawking,
  title     = {{H}awking Radiation of the {D}irac Field via an Anomalous Method},
  author    = {Jing, Jiliang},
  journal   = {Physical Review D},
  volume    = {70},
  number    = {6},
  pages     = {065004},
  year      = {2004},
  doi       = {10.1103/PhysRevD.70.065004}
}

@article{brill1957interaction,
  title     = {Interaction of Neutrinos and Gravitational Fields},
  author    = {Brill, Dieter R. and Wheeler, John A.},
  journal   = {Reviews of Modern Physics},
  volume    = {29},
  number    = {3},
  pages     = {465--479},
  year      = {1957},
  doi       = {10.1103/RevModPhys.29.465}
}

@article{damour1976black,
  title     = {Black-Hole Evaporation in the {K}lein--{S}auter--{H}eisenberg--{E}uler Formalism},
  author    = {Damour, T. and Ruffini, R.},
  journal   = {Physical Review D},
  volume    = {14},
  number    = {2},
  pages     = {332--334},
  year      = {1976},
  doi       = {10.1103/PhysRevD.14.332}
}

@book{birrell1984quantum,
  title     = {Quantum Fields in Curved Space},
  author    = {Birrell, N. D. and Davies, P. C. W.},
  publisher = {Cambridge University Press},
  year      = {1984},
  doi       = {10.1017/CBO9780511622632}
}

@article{wootters1998quantum,
  title     = {Entanglement of Formation of an Arbitrary State of Two Qubits},
  author    = {Wootters, William K.},
  journal   = {Physical Review Letters},
  volume    = {80},
  number    = {10},
  pages     = {2245--2248},
  year      = {1998},
  doi       = {10.1103/PhysRevLett.80.2245}
}

@article{hashemi2012genuinely,
  title     = {Genuinely Multipartite Concurrence of {$N$}-qubit {$X$}-matrices},
  author    = {Hashemi Rafsanjani, S. M. and Huber, M. and Broadbent, C. J. and Eberly, J. H.},
  journal   = {Physical Review A},
  volume    = {86},
  number    = {6},
  pages     = {062303},
  year      = {2012},
  doi       = {10.1103/PhysRevA.86.062303}
}

@article{horodecki1995violating,
  title     = {Violating {B}ell Inequality by Mixed Spin-{$\tfrac{1}{2}$} States: Necessary and Sufficient Condition},
  author    = {Horodecki, R. and Horodecki, P. and Horodecki, M.},
  journal   = {Physics Letters A},
  volume    = {200},
  number    = {5},
  pages     = {340--344},
  year      = {1995},
  doi       = {10.1016/0375-9601(95)00214-N}
}

@article{clauser1969proposed,
  title     = {Proposed Experiment to Test Local Hidden-Variable Theories},
  author    = {Clauser, John F. and Horne, Michael A. and Shimony, Abner and Holt, Richard A.},
  journal   = {Physical Review Letters},
  volume    = {23},
  number    = {15},
  pages     = {880--884},
  year      = {1969},
  doi       = {10.1103/PhysRevLett.23.880}
}

@article{adesso2007entanglement,
  title     = {Entanglement of Dirac fields in noninertial frames},
  author    = {Adesso, Gerardo and Fuentes-Schuller, Ivette and Ericsson, Marie},
  journal   = {Physical Review A},
  volume    = {76},
  number    = {6},
  pages     = {062112},
  year      = {2007},
  doi       = {10.1103/PhysRevA.76.062112}
}

@article{Ahmadi2014,
  title     = {Relativistic Quantum Metrology: Exploiting Relativity to Improve Quantum Measurement Technologies},
  author    = {Ahmadi, Mehdi and Bruschi, David Edward and Sab{\'\i}n, Carlos and Adesso, Gerardo and Fuentes, Ivette},
  journal   = {Scientific Reports},
  volume    = {4},
  pages     = {4996},
  year      = {2014},
  doi       = {10.1038/srep04996}
}

@article{Mann2012,
  title     = {Relativistic quantum information},
  author    = {Mann, Robert B. and Ralph, Timothy C.},
  journal   = {Classical and Quantum Gravity},
  volume    = {29},
  number    = {22},
  pages     = {220301},
  year      = {2012},
  doi       = {10.1088/0264-9381/29/22/220301}
}

@article{Peres2004,
  title     = {Quantum information and relativity theory},
  author    = {Peres, Asher and Terno, Daniel R.},
  journal   = {Reviews of Modern Physics},
  volume    = {76},
  number    = {1},
  pages     = {93--123},
  year      = {2004},
  doi       = {10.1103/RevModPhys.76.93}
}

@article{carballorubio2018phenomenological,
  title     = {Phenomenological aspects of black holes beyond general relativity},
  author    = {Carballo-Rubio, Raul and Di Filippo, Francesco and Liberati, Stefano and Visser, Matt},
  journal   = {Physical Review D},
  volume    = {98},
  number    = {12},
  pages     = {124009},
  year      = {2018},
  doi       = {10.1103/PhysRevD.98.124009}
}

@article{carballorubio2022inner,
  title     = {Inner horizon instability and the unstable cores of regular black holes},
  author    = {Carballo-Rubio, Raul and Di Filippo, Francesco and Liberati, Stefano and Pacilio, Costantino and Visser, Matt},
  journal   = {Journal of High Energy Physics},
  volume    = {2022},
  number    = {05},
  pages     = {132},
  year      = {2022},
  doi       = {10.1007/JHEP05(2022)132}
}

@article{kraus1983states,
  title     = {States, Effects, and Operations: Fundamental Notions of Quantum Theory},
  author    = {Kraus, Karl},
  journal   = {Lecture Notes in Physics},
  volume    = {190},
  publisher = {Springer},
  year      = {1983}
}

@article{unruh1976notes,
  title     = {Notes on black hole evaporation},
  author    = {Unruh, William G.},
  journal   = {Physical Review D},
  volume    = {14},
  number    = {4},
  pages     = {870--892},
  year      = {1976},
  doi       = {10.1103/PhysRevD.14.870}
}

@article{israel1976thermo,
  title     = {Thermo-field dynamics of black holes},
  author    = {Israel, Werner},
  journal   = {Physics Letters A},
  volume    = {57},
  number    = {2},
  pages     = {107--110},
  year      = {1976},
  doi       = {10.1016/0375-9601(76)90178-X}
}

@article{hartle1976path,
  title     = {Path-integral derivation of black-hole radiance},
  author    = {Hartle, James B. and Hawking, Stephen W.},
  journal   = {Physical Review D},
  volume    = {13},
  number    = {8},
  pages     = {2188--2203},
  year      = {1976},
  doi       = {10.1103/PhysRevD.13.2188}
}

@article{tsirelson1980quantum,
  title     = {Quantum generalizations of {B}ell's inequality},
  author    = {Cirel'son, B. S.},
  journal   = {Letters in Mathematical Physics},
  volume    = {4},
  number    = {2},
  pages     = {93--100},
  year      = {1980},
  doi       = {10.1007/BF00417500}
}

@article{Alsing2006,
  title     = {Entanglement of {D}irac fields in non-inertial frames},
  author    = {Alsing, Paul M. and Fuentes-Schuller, Ivette and Mann, Robert B. and Tessier, Tracey E.},
  journal   = {Physical Review A},
  volume    = {74},
  number    = {3},
  pages     = {032326},
  year      = {2006},
  doi       = {10.1103/PhysRevA.74.032326}
}

@article{martin2011fermionic,
  title     = {Fermionic entanglement that survives a black hole},
  author    = {Mart{\'{\i}}n-Mart{\'{\i}}nez, Eduardo and Le{\'o}n, Juan},
  journal   = {Physical Review A},
  volume    = {80},
  number    = {4},
  pages     = {042318},
  year      = {2009},
  doi       = {10.1103/PhysRevA.80.042318}
}

@book{Takagi1986,
  title     = {Vacuum noise and stress induced by uniform acceleration: Hawking--Unruh effect in {R}indler manifold of arbitrary dimension},
  author    = {Takagi, Shin},
  publisher = {Progress of Theoretical Physics Supplement},
  volume    = {88},
  pages     = {1--142},
  year      = {1986},
  doi       = {10.1143/PTPS.88.1}
}

@article{PanYu2008,
  title     = {Hawking radiation of {D}irac particles via tunneling from the {R}eissner--{N}ordstr{\"o}m black hole},
  author    = {Pan, Q. and Jing, J.},
  journal   = {Modern Physics Letters A},
  volume    = {23},
  number    = {1},
  pages     = {25--34},
  year      = {2008},
  doi       = {10.1142/S0217732308025966}
}

@article{Martin-Martinez2010,
  title     = {Entanglement in non-inertial frames: degradation and survival},
  author    = {Mart{\'{\i}}n-Mart{\'{\i}}nez, Eduardo and Fuentes, Ivette},
  journal   = {Physical Review A},
  volume    = {83},
  number    = {5},
  pages     = {052306},
  year      = {2011},
  doi       = {10.1103/PhysRevA.83.052306}
}
\bibliographystyle{unsrt}

\end{document}